%% file: document.tex
\documentclass{sig-alternate-05-2015}

\usepackage[]{inputenc}
\usepackage[T1]{fontenc}
\usepackage{placeins}

\usepackage{amsmath,amssymb}
\usepackage{amssymb}
\usepackage{mathtools}
\usepackage{amsfonts}
\usepackage{algorithmic}
\usepackage{courier}
\usepackage{graphicx}
\usepackage[usenames, dvipsnames]{color}
\usepackage[framemethod=TikZ]{mdframed}
\usepackage{fancyvrb}
\usepackage{relsize}

\usepackage{float}

\newenvironment{requirement}
{\vspace{0.05in}
 \begin{mdframed}[roundcorner=10pt,backgroundcolor=gray!20]}
{\end{mdframed}}

\begin{document}

\CopyrightYear{2016} 
\setcopyright{acmcopyright}
\conferenceinfo{FormaliSE'16,}{May 15 2016, Austin, TX, USA}
\isbn{978-1-4503-4159-2/16/05}\acmPrice{\$15.00}
\doi{http://dx.doi.org/10.1145/2897667.2897675}

\title{Towards Synthesis from Assume-Guarantee Contracts involving Infinite
Theories: A Preliminary Report}

\numberofauthors{3}
\author{
\alignauthor
	Andreas Katis\\
	   \affaddr{Department of Computer Science and Engineering}\\
       \affaddr{University of Minnesota}\\
       \affaddr{200 Union Street}\\
       \affaddr{Minneapolis, MN 55455,USA}\\
       \email{katis001@umn.edu}
\alignauthor
	Andrew Gacek\\
       \affaddr{Rockwell Collins\\ Advanced Technology Center}\\
       \affaddr{400 Collins Rd. NE}\\
       \affaddr{Cedar Rapids, IA, 52498, USA}\\
       \email{andrew.gacek@\\rockwellcollins.com}
\alignauthor
	Michael W. Whalen
	   \affaddr{Department of Computer Science and Engineering}\\
       \affaddr{University of Minnesota}\\
       \affaddr{200 Union Street}\\
       \affaddr{Minneapolis, MN 55455,USA}\\
       \email{whalen@cs.umn.edu}}

\maketitle

\begin{abstract}
In previous work, we have introduced a contract-based {\em realizability checking} 
algorithm for assume-guarantee contracts involving infinite theories, such as 
linear integer/real arithmetic and uninterpreted functions over infinite domains.  
This algorithm can determine whether or not it is possible to construct a 
realization (i.e. an implementation) of an assume-guarantee contract.  
The algorithm is similar to k-induction model checking, but involves the use of 
quantifiers to determine implementability.

While our work on realizability is inherently useful for {\em virtual integration} 
in determining whether it is possible for suppliers to build software that meets 
a contract, it also provides the foundations to solving the more challenging problem 
of component synthesis. In this paper, we provide an initial synthesis 
algorithm for assume-guarantee contracts involving 
infinite theories.  To do so, we take advantage of our realizability checking 
procedure and a skolemization solver for $\forall\exists$-formulas, called
AE-VAL.  We show that it is possible to immediately
adapt our existing algorithm towards synthesis by using this solver, using a demonstration example.  We then discuss challenges towards creating a more robust synthesis algorithm.

\end{abstract}

\input{introduction}
\input{preliminaries}

\input{synthesis}
\input{related}
\input{conclusion}

\vspace{+1cm}

\section{Acknowledgments}
This work was funded by DARPA and AFRL under contract 4504789784 (Secure Mathematically-Assured Composition of Control Models), and by NASA under contract NNA13AA21C (Compositional Verification of Flight Critical Systems), and by NSF under grant CNS-1035715 (Assuring the safety, security, and reliability of medical device cyber physical systems).
\bibliographystyle{IEEEtran}
\bibliography{document}

\end{document}

%% file: introduction.tex
\section{Introduction}
The problem of automated synthesis of reactive systems using from propositional specifications is a very well studied area of research~\cite{gulwani2010dimensions}. By definition, the problem of synthesis entails the discovery of efficient algorithms able to construct a candidate program that is guaranteed to comply with the predefined specification. Inevitably, the related work on synthesis has tackled several sub-problems, such as that of function and template synthesis, as well as the weaker problem regarding the implementability, or otherwise, realizability of the specification.

In a similar fashion, a collaboration between Rockwell Collins and
the University of Minnesota has focused on designing tools that provide
compositional proofs of correctness~\cite{NFM2012:CoGaMiWhLaLu,Whalen13:WhatHow:TwinPeaksIEEESoftware,hilt2013,QFCS15:backes}.
In the context of synthesis, we recently introduced a decision procedure for determine the {\em realizability} of contracts involving infinite theories such as linear integer/real arithmetic and/or uninterpreted functions that is checkable by any SMT solver that supports quantification~\cite{Katis15:Realizability}. Furthermore, in~\cite{Katis:machine} we formally proved the soundness of our checking algorithm using the Coq interactive theorem prover. The realizability checking procedure is now part of the AGREE reasoning framework~\cite{NFM2012:CoGaMiWhLaLu}, which supports compositional
assume-guarantee contract reasoning over system architectural models written in
AADL~\cite{SAE:AADL}.

While checking the realizability of contracts provided us with fruitful results
and insight in several case studies, it also worked as solid ground towards the
development of an automatic component synthesis procedure. The
most important obstacle initially, was the inability of the SMT solver to
handle higher-order quantification. Fortunately, interesting directions to
solving this problem have already surfaced, either by extending an SMT solver
with native synthesis capabilities\cite{reynoldscounterexample}, or by providing
external algorithms that reduce the problem by efficient quantifier elimination methods~\cite{fedyukovichae}.

The main contribution of this paper is the implementation of a
component synthesis algorithm for infinite theories, using specifications
expressed in assume-guarantee contracts.  The algorithm heavily relies on our previous implementation for realizability checking, but also takes advantage of a recently published skolemizer for $\forall\exists$-formulas, named
AE-VAL. The main idea in this implementation is to effectively extract a Skolem
relation that is essentially, a collection of strategies, that can directly lead to an implementation which is guaranteed to comply to the corresponding contract.

In Section~\ref{sec:preliminaries} we provide the necessary background
definitions from our previous work on realizability checking.
Section~\ref{sec:synthesis} presents our approach to solving the synthesis
problem for assume-guarantee contracts using theories. Finally, in
Section~\ref{sec:related work} we give a brief historical background on the
related research work on synthesis, and we report our conclusions and upcoming
future work in Section~\ref{sec:conclusion}.

%% file: preliminaries.tex
\section{Preliminaries}
\label{sec:preliminaries}

In the remainder of the paper, we endeavor to solve a synthesis problem for
assume-guarantee contracts involving infinite theories.  Formally, an
implementation is a set of valid initial states $I$ and transition relation $T$ that implements the contract.  In this section, we introduce the necessary formal machinery to talk about {\em realizations} of an assume-guarantee contract.

\subsection{Example}
\begin{figure}[H]
	\centering
	\includegraphics[width=0.45\textwidth,height=\textheight,keepaspectratio]{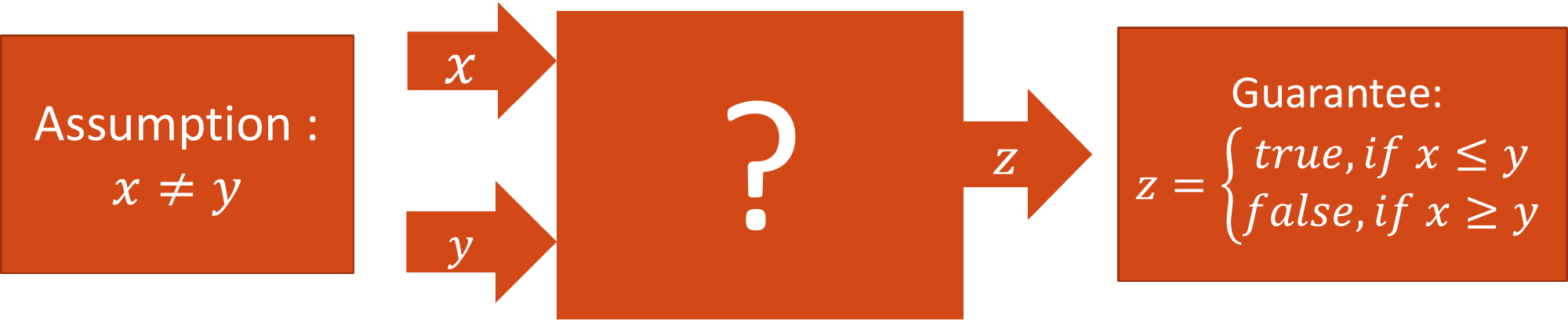}    	
	\caption{Example of a realizable contract}
	\label{fg:example}
\end{figure}

As an illustrative example, consider the contract specified in
Figure~\ref{fg:example}. The component to be designed consists of two
inputs, $x$ and $y$ and one output $z$. If we restrict our example to the case
of integer arithmetic, we can see that the contract assumes that the inputs will
never have the same value, and requires that the component's output is a Boolean
whose value depends on the comparison of the values of $x$ and $y$.

One can easily prove that an implementation able to satisfy the contract is
possible. The same though does not apply in the case where we omit the
assumption from our original contract. Given no constraints over the values that the inputs can take, we have a case where the implementation may
behave in an inconsistent manner regarding the value of $z$, thus making the
violation of the guarantees possible. In the first case, the contract is realizable, but in the second, we
cannot find an implementation that can provide us with an output that satisfies
the contract guarantees, for any valid input. These contracts are considered to
be unrealizable.

\subsection{Formal Definitions}
We use the types $state$ and $inputs$ to describe a state
and the set of inputs in the system, respectively. We define a
\textit{transition system} as a pair $(I,T)$, where $I$ is the set of initial states, of type $state \rightarrow
bool$ and $T$ is the transition relation, of type $state \rightarrow inputs
\rightarrow state \rightarrow bool$. 

A contract in this context is defined by its
\textit{assumptions} and \textit{guarantees}. The \textit{assumptions} $A$
impose constraints over the inputs, while the \textit{guarantees} are
further decomposed into the pair $(G_I,G_T)$ with $G_I$ describing the
valid initial states for the system, and $G_T$ specifying the new states to which
the system may transition, given a specific state and input. Note that
we do not necessarily
expect that a contract would be defined over all variables in the
transition system,
but we do not make any distinction between internal state variables
and outputs in the formalism.  This way, we can use state variables to, in some cases, simplify statements of guarantees.

Given the above, we expressed realizability as follows. A state $s$ is
\textit{viable} if, starting from $s$, the transition system is capable of continuously responding to
incoming valid inputs. Alternatively, $s$ is $viable$ if the transitional
guarantee $G_T$ infinitely holds, given valid inputs. As such,
viability is defined coinductively:
\begin{equation}
\label{eq:viable}
Viable(s) = \forall i. A(s,i) \Rightarrow \exists s'. G_{T}(s,i,s') \wedge
Viable(s')
\end{equation}
Using the definition of $viable$, a
contract is $realizable$ if and only if
\begin{equation}
\label{eq:realizable}
 \exists s.~ G_I(s) \land Viable(s).
\end{equation}

To use this coinductive formula in a model checking framework, we had to further
massage the definition into one that resembles the principle of
\textit{k-induction}. As such, the base step of the induction ensures that
given a state, $G_T$ can keep responding to valid inputs for at least $n$ steps
in the future. We call this state \textit{finitely viable}, written
$Viable_{n}(s)$:
\begin{multline}
\label{ml:fv}
		\forall i_1. A(s,i_1) \Rightarrow \exists s_1. G_T(s,i_1,s_1) \wedge~ \\
			\hspace{+0.5cm} \forall i_2. A(s_1,i_2) \Rightarrow \exists s_2.
			 G_T(s_1,i_2,s_2) \wedge \ldots \wedge~ \\
			 \hspace{+1.5cm} \forall i_n. A(s_{n-1},i_n) \Rightarrow \exists s_n. G_T(s_{n-1},i_n,s_n)
	\end{multline}

On the other hand, the inductive step checks whether a path starting from a
finitely viable state can be further extended one-step. We call states that
build such paths \textit{extendable}, written $Extend_{n}(s)$:
	\begin{multline}
	\label{ml:extendable}
		\forall i_1,s_1,\ldots,i_n,s_n. \\
		\hspace{-2cm} A(s,i_1) \wedge G_T(s,i_1,s_1) \wedge \ldots \wedge \\
		A(s_{n-1}, i_n) \wedge G_T(s_{n-1},i_n,s_n) \Rightarrow \\
		\hspace{+2cm} \forall i. A(s_n,i) \Rightarrow \exists s'. G_T(s_n,i,s')
	\end{multline}
	
Considering these underapproximations, the algorithm is split into two separate
checks. For the \textit{BaseCheck}, we ensure that there exists an initial
finitely viable state,
\begin{equation}
\label{bcheck}
BaseCheck(n) = \exists s. G_I(s) \wedge Viable_n(s)
\end{equation}
while the \textit{ExtendCheck} tries to prove that all
valid states are extendable.
\begin{equation}
\label{eq:echeck}
ExtendCheck(n) = \forall s. Extend_n(s)
\end{equation}
Due to the definition of finite viability containing $2n$
quantifier alternations, we can not practically use \textit{BaseCheck} as the
current SMT solvers struggle solving formulas of such structure. Therefore we
finally proposed a simplified version of \textit{BaseCheck}, which essentially
tries to prove that all initial states are extendable for any $k \leq n$.
\begin{equation}
\label{eq:sbcheck}
BaseCheck'(n) = \forall k \leq n. (\forall s. G_I(s)
	  	\Rightarrow Extend_k(s))
\end{equation}

Even though the simplified definition is more simple for an SMT solver to
process, it comes with a cost, as it introduces cases of realizable contracts
that are considered to be unrealizable from the algorithm. To the extent of our
experiments, such a case has yet to be met, as it inherently requires the user
to purposely define contracts of such behavior.

%% file: synthesis.tex
\section{Synthesis from Contracts}
\label{sec:synthesis}

With a sound implementation of the realizability checking algorithm at
hand, the next step was to tackle the more
interesting problem of synthesis, i.e. the automated derivation of
implementations that would be safe in terms of satisfying the constraints
defined by the component's contract. The intuition behind solving the
synthesis problem in our context relies on finding a set of initial states
$I$ and a transition relation $T$ that would satisfy the requirements
specified in the contract. Unfortunately, the lack of power in SMT solvers in
terms of solving formulas that contain higher-order quantification immediately
ruled out the prospect of using one as our primary synthesis tool. Therefore, an alternative work from Fedyukovich et al.~\cite{fedyukovichae,fedyukovich2014automated} on a skolemizer for $\forall\exists$-formulas on linear integer arithmetic was chosen to be used as a means of extracting a witness that could directly be used in component synthesis.

The tool, called AE-VAL is using the Model-Based Projection technique
in~\cite{komuravelli2014smt} to validate $\forall\exists$-formulas, based on
Loos-Weispfenning quantifier elimination~\cite{loos1993applying}.
As part of the procedure, a Skolem relation is provided for the existentially quantified variables of the formula. The algorithm
initially distributes the models of the original formula into disjoint
uninterpreted partitions, with a local Skolem relation being computed for each
partition in the process. From there, the use of a Horn-solver provides an
interpretation for each partition, and a final global Skolem relation is produced.

The idea behind our approach to solving the synthesis problem is simple.
Consider the checks~\ref{eq:sbcheck} and~\ref{eq:echeck} that the realizability
checking algorithm is using. \textit{BaseCheck'} is still
necessary for the synthesis problem to ensure that all initial states in the problem are valid.
\textit{ExtendCheck} on the other hand can be further used in actually
synthesizing implementations. The check tries to prove that every valid state in
our system is extendable, i.e. all states can be starting points to paths that
comply to the system contract, and furthermore are extendable by one step:
	\begin{multline*}
		\forall i_1,s_1,\ldots,i_n,s_n. \\
		\hspace{-2cm} A(s,i_1) \wedge G_T(s,i_1,s_1) \wedge \ldots \wedge \\
		A(s_{n-1}, i_n) \wedge G_T(s_{n-1},i_n,s_n) \Rightarrow \\
		\hspace{+2cm} \forall i. A(s_n,i) \Rightarrow \exists s'. G_T(s_n,i,s')
	\end{multline*}
which can be rewritten:
	\begin{multline}
	\label{ml:extendable2}
		\forall i_1,s_1,\ldots,i_n,s_n,i. \\
		\hspace{-2cm} A(s,i_1) \wedge G_T(s,i_1,s_1) \wedge \ldots \wedge \\
		A(s_{n-1}, i_n) \wedge G_T(s_{n-1},i_n,s_n) \wedge A(s_n,i) \Rightarrow \\
		\hspace{+2cm} \exists s'. G_T(s_n,i,s')
	\end{multline}
Such a formula is exactly what is required by a $\forall\exists$
solver such as AE-VAL in order to produce the witness for the existential variables $s'$.
AE-VAL solves this formula by providing an assignment for each existential variable in a piecewise relation based on a partitioning based on assignments to the universal variables.  In other words, by examining a bounded history of the state and input variable values in the contract (the universally quantified variables in Formula~\ref{ml:extendable2}), we determine the next values of the state variables.  An example of a portion of this partitioning is shown in Figure~\ref{fig:skolem-rel}.
In other words, the Skolem relation contains, starting from a valid
initial state of variables, strategies in terms of how the new state is
selected, in such a way, that the contract is not violated.

\begin{figure}
\begin{small}
\begin{verbatim}
// for each variable in I or S,
//   create an array of size k.
//   then initialize initial state values
assign_GI_witness_to_S;
update_array_history;

// Perform bounded 'base check' synthesis
read_inputs;
base_check'_1_solution;
update_array_history;
...
read_inputs;
base_check'_k_solution;
update_array_history;

// Perform recurrence from 'extends' check
while(1) {
 read_inputs;
 extend_check_k_solution;
 update_array_history;
}
\end{verbatim}
\end{small}
\caption{Algorithm skeleton for synthesis}
\label{fig:algorithm}
\end{figure}

\noindent

Thus, we can construct the skeleton of an algorithm as shown in Figure~\ref{fig:algorithm}.  
We begin by creating an array for each input and history variable up to depth
$k$, where $k$ is the depth at which we found a solution to our realizability algorithm.
In each array, the zeroth element is the `current' value of the variable, the first element is the previous value, and the $(k-1)$'th value is the $(k-1)$-step previous value.
We then generate witnesses for each of the {\em BaseCheck'} instances of
successive depth using the AE-VAL solver to describe the initial behavior of the
implementation up to depth $k$.  This process starts from the memory-free
description of the initial state ($G_I$).  There are two `helper' operations:
{\em update\_array\_history} shifts each array's elements one position forward
(the $(k-1)$'th value is simply forgotten), and {\em read\_inputs} reads the current values of inputs into the zeroth element of the input variable arrays.  Once the history is entirely initialized using the {\em BaseCheck'} witness values, we enter a recurrence loop where we use the solution of the {\em ExtendCheck} to describe the next value of outputs.

\subsection{Synthesis Example}
As an example that demonstrates the process, consider the contract created for
a mode controller in a simple microwave model of 260 lines of code that was
used as one of the base case studies in~\cite{Katis15:Realizability}. The
controller has four inputs, \textit{start} which is used to indicate whether the microwave is at an initial
state or not, \textit{clear} that is used as a stop signal for the system,
\textit{seconds\_to\_cook} as a countdown timer and \textit{door\_closed} as an
indicator that the microwave's door is closed or not. The controller returns the
current state of the microwave's mode using \textit{cooking\_mode}. The contract
consists of one assumption and nine guarantees, which are shown below
informally, as well as formally in AADL. A library named \textit{defs} is used
to define auxiliary functions, such as \textit{rising\_edge()} which returns ``true'' when the corresponding signal is at its rising edge, and \textit{initially\_true()} which is used to
check a variable's value at the component's initial state.

\begin{requirement}
\textbf{MC Assumption} -- seconds\_to\_cook is greater than or equal to
zero.
\begin{verbatim}
seconds_to_cook >= 0;
\end{verbatim}
\label{req:micasm}
\end{requirement}
\begin{requirement}
\textbf{MC Guarantee-0} -- The range of the cooking\_mode variable shall be
[1..3].
\begin{verbatim}
cooking_mode >= 1 and cooking_mode <= 3;
\end{verbatim}
\label{req:mic0}
\end{requirement}
\begin{requirement}
\textbf{MC Guarantee-1} -- The microwave shall be in cooking mode only when the
door is closed.
\begin{Verbatim}[obeytabs,fontsize=\small]
is_running => door_closed;
\end{Verbatim}
\label{req:mic1}
\end{requirement}

\begin{requirement}
\textbf{MC Guarantee-2} -- The microwave shall be in setup mode in the initial
state.
\begin{verbatim}
(defs.initially_true(start)) => is_setup;
\end{verbatim}
\label{req:mic2}
\end{requirement}
\begin{requirement}
\textbf{MC Guarantee-3} -- At the instant the microwave starts running, it shall
be in the cooking mode if the door is closed.
\begin{Verbatim}[obeytabs,fontsize=]
(defs.rising_edge(is_running) and
	       door_closed) => is_cooking;
\end{Verbatim}
\label{req:mic3}
\end{requirement}
\begin{requirement}
\textbf{MC Guarantee-4} -- At the instant the microwave starts running, it shall
enter the suspended mode if the door is open.
\begin{Verbatim}[obeytabs,fontsize=]
(defs.rising_edge(is_running) and
	not door_closed) => is_suspended;
\end{Verbatim}
\label{req:mic4}
\end{requirement}
\begin{requirement}
\textbf{MC Guarantee-5} -- At the instant the clear button is pressed, if the
microwave was cooking, then the microwave shall stop cooking.
\begin{Verbatim}[obeytabs,fontsize=]
(defs.rising_edge(clear) and
	is_cooking) => not is_cooking;
\end{Verbatim}
\label{req:mic5}
\end{requirement}
\begin{requirement}
\textbf{MC Guarantee-6} -- At the instant when the clear button is pressed, if
the microwave is in suspended mode, it shall enter the setup mode.
\begin{Verbatim}[obeytabs,fontsize=]
(defs.rising_edge(clear) and
	is_suspended) => is_setup;
\end{Verbatim}
\label{req:mic6}
\end{requirement}
\begin{requirement}
\textbf{MC Guarantee-7} -- If suspended, at the instant the start key is pressed
the microwave shall enter cooking mode if the door is closed.
\begin{Verbatim}[obeytabs]
(defs.rising_edge(start) and is_suspended
	and door_closed) => is_cooking;
\end{Verbatim}
\label{req:mic7}
\end{requirement}
\begin{requirement}
\textbf{MC Guarantee-8} -- If seconds\_to\_cook = 0, microwave will be in setup
mode.
\begin{Verbatim}[obeytabs,fontsize=]
(seconds_to_cook = 0) => is_setup;
\end{Verbatim}
\label{req:mic8}
\end{requirement}

The contract is then translated from AADL into an equivalent Lustre program that
is then given as input to the JKind model checker~\cite{jkind}, where our
realizability algorithm is implemented as a separate feature. From JKind, the
Lustre specification is further translated into the SMT-LIB v2 format. In
reality, the program is split into two different processes that run in parallel and correspond to the
checks~\ref{eq:echeck} and~\ref{eq:sbcheck} that our synthesis algorithm is
using. Considering the fact that the
contract is realizable, we impose the negation of our target
$\forall\exists$-formula as a query to the AE-VAL skolemizer during the last
step of \textit{ExtendCheck}. AE-VAL responds that the original formula can be
satisfied, and provides a Skolem relation, a part of which is shown in Figure~\ref{fig:skolem-rel}.

\begin{figure}
\begin{Verbatim}[obeytabs,fontsize=\tiny]
ite([&&
    $defs__rising_edge~1.Mode_Control_Impl_Instance__signal$0
    !($Mode_Control_Impl_Instance__seconds_to_cook$0>=0)
    !$defs__initially_true~0.Mode_Control_Impl_Instance__result$0
  ], [&&
    $Mode_Control_Impl_Instance__is_setup$0
    $defs__rising_edge~1.Mode_Control_Impl_Instance__re$0
    !$Mode_Control_Impl_Instance__is_cooking$0
    $defs__rising_edge~1.Mode_Control_Impl_Instance__signal$0
    !$_TOTAL_COMP_HIST$0
    !$_SYSTEM_ASSUMP_HIST$0
    !$Mode_Control_Impl_Instance__is_suspended$0
    !$Mode_Control_Impl_Instance__is_running$0
    !$defs__rising_edge~0.Mode_Control_Impl_Instance__re$0
    !$defs__initially_true~0.Mode_Control_Impl_Instance__b$0
    !$defs__initially_true~0.Mode_Control_Impl_Instance__result$0
    !$defs__rising_edge~2.Mode_Control_Impl_Instance__re$0
    !$defs__rising_edge~2.Mode_Control_Impl_Instance__signal$0
  ], ite([&&
      %init
      $_SYS_GUARANTEE_2$0
      !($Mode_Control_Impl_Instance__seconds_to_cook$0>=0)
      !$defs__rising_edge~1.Mode_Control_Impl_Instance__signal$0
      !$defs__initially_true~0.Mode_Control_Impl_Instance__b$0
    ], ...))
\end{Verbatim}
\caption{A portion of the Skolem relation generated for the Microwave Mode Logic}
\label{fig:skolem-rel}
\end{figure}
As seen in Figure~\ref{fig:skolem-rel}, the Skolem relation is composed of nested
\textit{if-then-else} blocks, which indicate the possible valid transitions the
implementation can follow given a specific state, without
violating the contract. Each variable appearing in the conjuncts of
the relation is named uniquely after the state that it refers to, using the
\textit{\$X} postfix, where $X$ is an integer. In addition to each state's
variables, we keep track of whether each state is initial using the variable
\textit{\%init}.

The structure of the Skolem relation is simple enough to translate
into a program in a mainstream language.
We need implementations that are able to keep track of the
current state variables, the current inputs, as well as some history
about the variable values in previous states. This can easily be handled, for
example, in C with the use of arrays to keep record of each variable's $k$ last
values, and the use of functions that update each variable's corresponding array
to reflect the changes following a new step using the transition relation.

%% file: related.tex
\section{related work}
\label{sec:related work}

The problem of program synthesis was first expressed formally in the early
1970s~\cite{manna1971toward} as a potentially important area of study and
research. 
Pnuelli and Rosner use the term
$\textit{implementability}$ in~\cite{Pnueli89} to refer to the problem of synthesis for
propositional LTL. Additionally, the authors in~\cite{Pnueli89} proved that the
lower-bound time complexity of the problem is doubly exponential, in the worst
case. In the following years, several techniques were introduced to deal with
the synthesis problem in a more efficient way for subsets of propositional LTL
\cite{Klein10}, simple LTL formulas (\cite{Bohy12}, \cite{Tini03}), as well as
in a component-based approach \cite{Chatterjee07} and specifications based on
other temporal logics (\cite{benevs2012factorization}, \cite{Hamza10}), such as SIS
\cite{Aziz95}.

In 2010, a survey from Sumit Gulwani described the 
directions that future research will focus on, towards the
road of fully automated synthesis of programs~\cite{gulwani2010dimensions}.
The approaches that have been proposed are many, and differ on many aspects,
either in terms of the specifications that are being exercised, or the reasoning
behind the synthesis algorithm itself. On the one hand, template-based
synthesis~\cite{srivastava2013template} is focused on the exploration of
programs that satisfy a specification that is refined after each
iteration, following the basic principles of deductive synthesis. Inductive
synthesis, on the other hand, is an active area of research where the main goal
is the generation of an inductive invariant that can be used to describe the
space of programs that are guaranteed to satisfy the given
specification~\cite{flener2001inductive}.
This idea is mainly supported by the use of SMT solvers to guide the invariant
refinement through traces that violate the requirements, known as
counterexamples. Recently published work on extending SMT solvers with
counterexample-guided synthesis shows that they can eventually be
used as an alternative to solving the problem under certain domains of
arithmetic~\cite{reynoldscounterexample}.

Finally, an interesting and relevant work has been done regarding the solution
to the controllability problem using in \cite{micheli_aaai_2012}
\cite{micheli_cp_2012} and \cite{micheli_constraints_2014}, which involves the
decision on the existence a strategy that assigns certain values to a set of
controllable activities, with respect to a set of uncontrollable ones.

Our approach relies on the idea of extracting programs that satisfy the
constraints from the proof of their realizability that is produced by a
sophisticated theorem prover. The proof itself is provided
through a model checking approach that follows the k-induction principle. To the
best of our knowledge this is the first attempt on providing a synthesis
algorithm for an assume-guarantee framework, using infinite theories.

%% file: conclusion.tex
\section{conclusion}
\label{sec:conclusion}

In this paper we present the first known algorithm of implementation synthesis
from assume-guarantee contracts, using theories. To achieve this, we took
advantage of our previous work, by extracting programs directly from the
contract's proof of realizability. Additionally, the algorithm depends on the
extraction of Skolem relations from the AE-VAL decision procedure for
$\forall\exists$-formulas.

Future work involves exploring the solution to many obstacles that stand still.
First, we want to aim towards extending our current approach to other theories
like linear real arithmetic, as AE-VAL currently only supports integer
arithmetic. Another goal that we are interested in exploring is the definition
of a better realizability checking algorithm based on the idea of invariant
generation, using the idea of property directed
reachability~\cite{bradley11,cimatti2014ic3,een2011efficient}. Another problem
to potentially consider are cases where the provided implementation cannot
actually be used in practice. This is an interesting area of research
due to the use of infinite theories in our approach, which may result in
implementations that use infinite precision, a feature that cannot be
practically achieved by any real program.

Several other directions to improving our existing synthesis algorithm involve
the improvement of representations in our context. For example the transition
relation often takes up a big portion of the final SMT-LIB output that is given
to AE-VAL to process, and is relatively hard to process. The same applies to the
Skolem relation, which for this example is almost 900 lines of nested
\textit{if-then-else} blocks. An interesting approach to improving the algorithm's performance relies in the translation
of the original data-flow program from Lustre to a finite state machine, using a
sophisticated compilation methods as the ones presented
in~\cite{Halbwachs91:codegen}.
The disadvantage of using this approach is mainly that the final state machine
is not guaranteed to be minimal, due to the declarative nature of the programs
that we exercise. As a final remark, we intend to formally verify the synthesis
algorithm presented in this paper, by extending the proof that has already been
constructed for our algorithm on realizability checking.